\documentclass{basi}
\usepackage{epsfig}
\usepackage{txfonts}
\begin{document}
\title[ The GMRT study  of extremely faint dwarf irregular galaxies]{GMRT study of extremely faint dwarf irregular galaxies}
\author[Begum \& Chengalur]
       {A. Begum,$^1$\thanks{e-mail:ayesha@ncra.tifr.res.in} and
       Jayaram N. Chengalur,$^1$ \\
	$^1$National Centre for Radio Astrophysics, Post Bag 3, Ganeshkhind, Pune 411 007
        }
\maketitle
\label{firstpage}
\begin{abstract}
We present results of deep, high velocity resolution ($\sim$1.6 kms$^{-1}$) Giant Meterwave
Radio Telescope (GMRT) HI 21 cm-line observations of  a sample of extremely faint 
(M$_B>-$12.5) dwarf irregular galaxies. Unlike previous studies, we find large scale velocity gradients 
in even the faintest galaxies of our sample i.e. their velocity fields are not 
``chaotic". For some of our sample galaxies, where the velocity fields are in 
fact completely consistent with ordered rotation, we derive rotation curves 
and determine the structure of the dark matter halos of the galaxies from mass 
modeling, using the ``asymmetric drift" corrected rotation velocities. Finally, we compile
from literature a sample of galaxies with known dark matter distribution,
estimated from HI synthesis observations, and discuss the results and their 
implications for hierarchical galaxy formation models. 
 
\end{abstract}

\section{Introduction}
\label{sec:intro}

In hierarchical galaxy formation models, based on the cold dark matter (CDM) paradigm, 
very small dwarf galaxies are excellent candidates for earliest formed galaxies. Numerical 
simulations based on these models  make definitive predictions about the average density 
and shape of the dark matter distribution in  such galaxies. These dwarf 
systems are typically dark matter dominated, even in the innermost  regions, 
unlike bright galaxies where the stellar population is dynamically important. 
Dwarf galaxies could hence provide an unique opportunity to compare observations  
with the CDM simulations.

 According to the numerical  simulations, mass-density distribution 
in the inner  parts of the simulated dark matter halos could be well 
described by  a  cusp i.e.  r$^{-1}$ power law. This  cusp in the density 
distribution  manifests  itself as a  steeply rising rotation curve in the 
inner regions of galaxies. However, this prediction from CDM simulations 
is found to disagree with the observed rotation curves of several 
dwarf galaxies (e.g. deBlok \& Bosma 2002); the data indicate a constant 
density core dark matter distribution. Apart from the shape of the dark 
matter halos, numerical simulations also predict an anti-correlation 
between the characteristic density and the virial mass of the dark matter 
halos. In hierarchical scenarios, the  low mass halos form at the earlier times, when the 
background was higher, hence dwarf  galaxies  should have larger dark 
matter densities.

 All these  predictions of the numerical simulations were however untested 
in the faintest dwarf galaxies, as it was widely believed that very faint dwarf 
irregular galaxies do not show any systematic rotational motions. 
From a systematic study of the kinematics of a sample of dwarfs, C\^{o}t\'{e}, Carignan
\& Freeman  (2000)  found that normal rotation is  seen only in galaxies
brighter than -14 mag, while fainter dwarfs have disturbed kinematics.
This result is consistent with the earlier findings of Lo, Sargent \& Young (1993), who
from a study of a sample of dwarfs  (with M$_{B\rm} \sim -9$ to M$_{B\rm} \sim -15$)
found that very faint dwarf irregulars have chaotic velocity fields. However, most of the
previous studies were done with a coarser velocity resolution and modest sensitivities.

Inorder to study the kinematics of very faint dwarf irregular galaxies, we obtained 
high velocity resolution  and high sensitivity  GMRT HI 21 cm-line  observations of a sample 
of galaxies with M$_B>-$12.5. 
The GMRT has a hybrid configuration which simultaneously 
provides both high angular resolution ($\sim$ 2$''$ if one uses baselines between the arm antennas) 
as well as sensitivity to  extended emission (from baselines between the antennas in the central 
square). This unique hybrid configuration of the GMRT makes it an excellent facility 
to study such galaxies. The velocity resolution used for our observations was $\sim1.6$ kms$^{-1}$
and the  typical integration time on each source was $\sim 16~-$18 hrs, which gave  a  typical rms 
of $\sim$ 1.0 mJy/Beam per channel. The galaxies in our sample have typical HI masses $\sim$ 10
$^{6-8}$ M$_\odot$.

We present here the results obtained from our GMRT observations. 

\section{Results of The GMRT Observations}
\label{sec:result}

\begin{figure}[h!]
\rotatebox{-90}{\epsfig{file=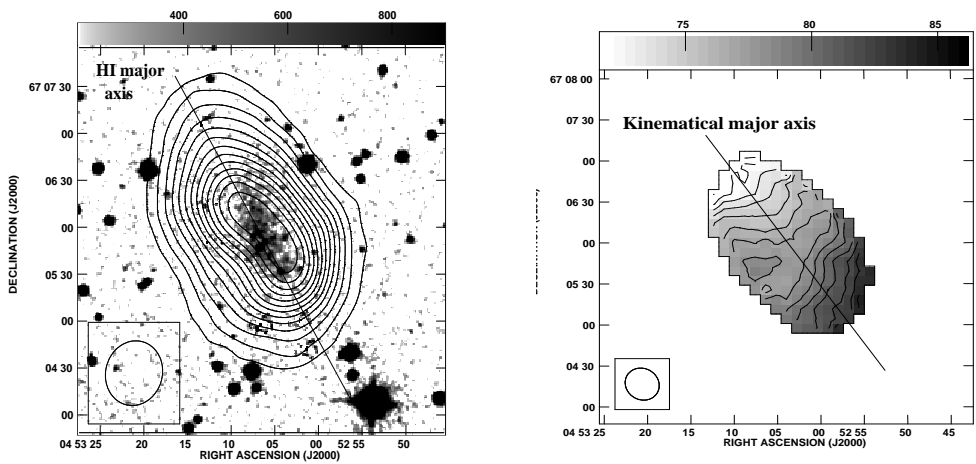,width=2.5in, height=5.5in}}
\caption{{\bf{[A]}}Integrated HI emission map of Cam B at
         $40^{''} \times 38^{''}$ resolution overlayed on the Digitised Sky Survey 
	Image. The contour levels are $3.7, 8.8, 19.1, 24.3, 29.4, 34.6, 39.8, 44.9, 50.1, 
	55.2, 60.4, 65.5~\&~ 70.7\times 10^{19}$ atoms~cm$^{-2}$.          
{\bf{[B]}}The HI velocity field of Cam~B at $24^{''}\times 22^{''}$ 
          resolution. The contours are in steps of 1~kms$^{-1}$ and
          range from 70.0~kms$^{-1}$ (the extreme North East contour)
          to 84.0~kms$^{-1}$ (the extreme South West contour).
         }
\label{fig:camb_mom}
\end{figure}

Figs.1[B],3[B] show the velocity fields for some of the galaxies in our sample.
Our GMRT observations clearly  show that the earlier conclusions about the kinematics
of faintest galaxies were a consequence of observational bias.  Since most of the 
earlier  studies of faint dwarf galaxies were done with coarser  
velocity resolutions ($\sim 6 - 7$ kms$^{-1}$) and modest sensitivities,
hence, such observations could not  detect systematic gradients, which 
are typically $\sim 10-15$ kms$^{-1}$, in the velocity fields of such faint 
dwarfs. On the other hand, our high velocity resolution and high 
sensitivity observations found large scale systematic velocity gradients 
in even the faintest galaxy in  our sample i.e.  DDO210. The velocity field for 
 DDO210, differs significantly from the  velocity field derived earlier by 
Lo et al.(1993). The pattern seen in the  GMRT velocity field of DDO210 (fig.~\ref{fig:camb_mom}[B]) is, 
to zeroth order, consistent with that expected from rotation, on the other 
hand, the velocity field derived by Lo et al.(1993) based on a coarser  velocity 
resolution of $\sim$ 6 kms$^{-1}$, is indeed ``chaotic". This difference in the 
observed kinematics suggests that high velocity resolution and high sensitivity 
is  crucial in determining the systematic gradients in the velocity fields of 
faint galaxies. For some of the galaxies in our sample this large scale 
systematic gradients could be modeled as systematic rotation, hence allowing 
us to derive the rotation curve for  those galaxies and to determine the structure 
of their dark matter halos through mass modelling. The rest of this  paper discusses 
the results obtained from the detailed mass modelling of two of our 
sample galaxies viz. Camelopardalis B (Cam B;M$_B\sim-$10.9) and DDO210 (M$_B\sim-$10.6).

\subsection{Camelopardalis B}
\label{ssec:camb}

Fig.~\ref{fig:camb_mom}[A] shows the integrated  HI column density image of Cam B at 40$''\times 38''$
resolution overlayed on the optical DSS image. The HI mass obtained from the integrated profile
(taking the distance to the galaxy to be 2.2 Mpc) is 5.3$\pm$0.5 $\times10^6$M$_\odot$ and the M$_{HI}$/L$_B$
ratio is found to be 1.4 in solar units.
 
 The velocity field derived from the moment analysis of 24$''\times 22''$ resolution data is shown
in the Fig.~\ref{fig:camb_mom}[B]. The velocity field is regular and the isovelocity contours are approximately
parallel, this is a signature of rigid body rotation. The kinematical major axis of the galaxy has a position
angle $\sim$ 215$^\circ$, i.e. it is well aligned with the major axis of both the HI distribution and the optical 
emission. 

\begin{figure}[h!]
\rotatebox{-90}{\epsfig{file=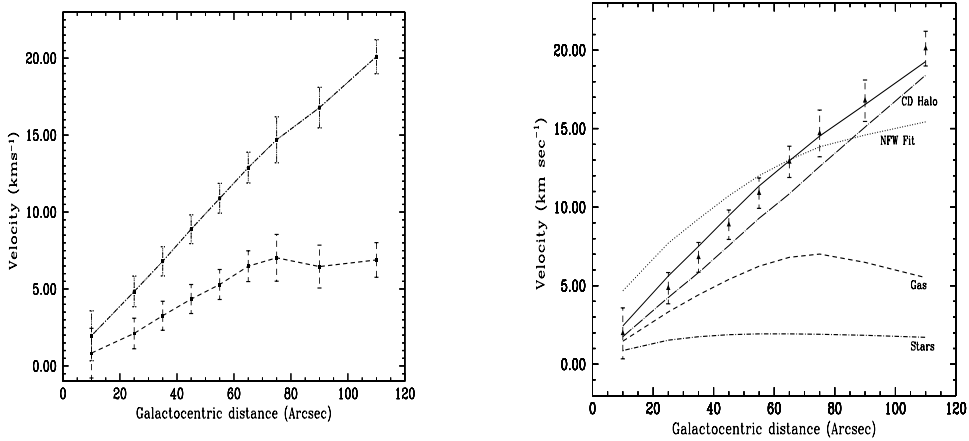,width=3.0in,height=5.5in}}
\caption{{\bf{[A]}}The derived rotation curve for Cam B (dashes)
and the rotation curve after applying
          the asymmetric drift correction (dash dots).
{\bf{[B]}}Mass models for Cam B using the corrected rotation curve.
The points are the observed data. The total mass of gaseous disk (dashed line)
is $6.6\times10^6 M_\odot$ (after scaling the total HI mass by a factor of 1.25, to include
the contibution of primordial Helium). The stellar disk (short dash dot line) has
$\Upsilon_V=0.2$, giving a stellar mass of $0.7 \times10^6 M_\odot$. The
best fit total rotation curve for the constant density halo model is shown as
a solid line, while the contribution of the halo itself is shown as a
long dash dot line (the halo density is $\rho_0=13.7\times10^{-3}
M_\odot$ pc$^{-3}$). The best fit total rotation curve for an NFW type halo
(for $c=1.0$ and $\Upsilon_V=0.0$) is shown as a dotted line.
         }
\label{fig:camb_vrot}
\end{figure}

       The rotation curve for Cam B was derived  from the velocity fields at 40$''\times 38''$, 24$''\times 22''$
and 16$''\times 14''$ resolution using the tilted ring method (see Begum, Chengalur and Hopp 2003 for details). The 
derived hybrid rotation curve is shown as a dashed curve in Fig.~\ref{fig:camb_vrot}[A]. We find that the peak 
inclination corrected rotational velocity  for Cam B ($\sim$ 7.0 kms$^{-1}$) is comparable to the observed HI
dispersion i.e. V$_{\rm{max}}/\sigma_{\rm{HI}}\sim1.0$. This implies that the random motions provide significant 
dynamical support to the system. In other words, the observed rotational velocities underestimate the total 
dynamical mass in the galaxy due to a significant pressure support of  the HI gas. Hence, 
 the observed rotation velocities were corrected for this pressure support, using the usual 
``asymmetric drift" correction, before constructing mass models (see Begum et al (2003) for details). 
The dot-dashed curve in Fig.~\ref{fig:camb_vrot}[A] shows the ``asymmetric drift" corrected rotation curve.

Using the ``asymmetric drift"  corrected curve, mass models for Cam B were derived.
The best fit mass model using a modified isothermal halo is shown 
in Fig.~\ref{fig:camb_vrot}[B]. Also shown in the figure is
the best fit total rotation curve for an NFW type halo.
As can be seen, the kinematics of Cam~B is  well fit with
a modified isothermal halo while an NFW halo provides
a poor fit to the data.

 The ``asymmetric drift" corrected rotation curve for Cam B is
rising till the last measured point (Fig.~\ref{fig:camb_vrot}[B]), hence the core
radius of the isothermal halo could not be constrained
from the data. The best fit model for a constant density halo
gives central halo density ($\rho_0$) of 12.0$\times 10^{-3}~M_
\odot$~pc$^{-3}$.
The derived $\rho_0$ is relatively insensitive to the assumed
mass-to-light ratio of the stellar disk ($\Upsilon_V$). We found that
by changing  $\Upsilon_V$ from a value of 0 (minimum disk fit)
to a value of 2.0 (maximum disk fit), $\rho_0$ changes by $<$20\%.
  From the last measured point of the  observed rotation curve, a total dynamical mass of
1.1$\times 10^8 M_\odot$ is derived, i.e. at the last measured point
more than 90\% of the mass  of Cam~B is dark. Futher, the dominance of the
dark matter halo together with the linear shape of the rotation
curve (after correction for ``asymmetric drift'') means that one
cannot obtain a good fit to the rotation curve using an NFW halo
regardless of the assumed $\Upsilon_V$.

\begin{figure}[h!]
\rotatebox{0}{\epsfig{file=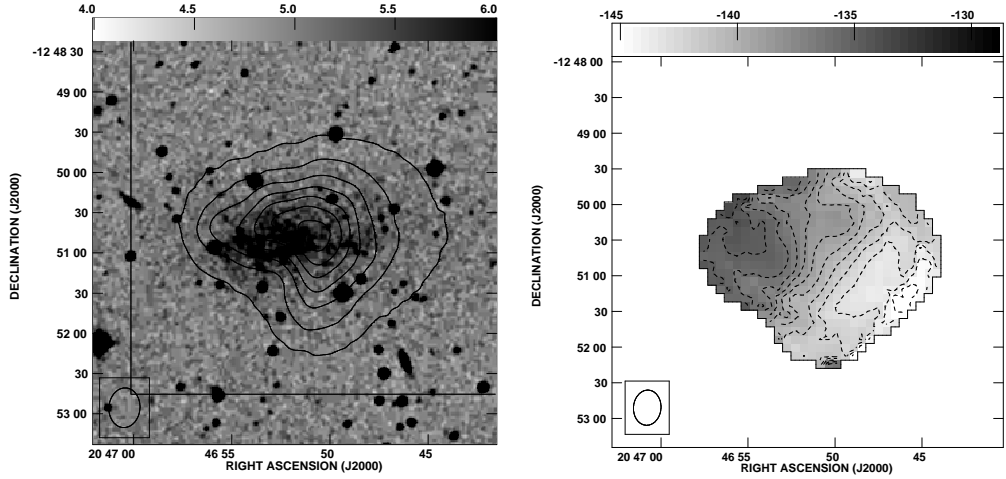,width=5.5in}}
\caption{{\bf{[A]}} The optical DSS image of DDO210 (greyscales) with
         the GMRT 44$^{''}\times37^{''}$  resolution integrated HI
         emission map (contours) overlayed. 
	The contour levels are $0.7, 10.2, 19.8, 29.3, 38.3, 45.5, 57.1, 
	67.5, 77.1, 86.7, 96.2, 105.7, 121.4~ \& ~124.8 \times 10^{19}$ atoms~cm$^{-2}$.
{\bf{[B]}}The HI velocity field of DDO210 at 29$^{''}\times 23^{''}$
          resolution. The contours are in steps of 1~kms$^{-1}$ and
          range from $-$145.0~kms$^{-1}$ to $-$133.0~kms$^{-1}$.
         }
\label{fig:ddo210_mom}
\end{figure}

\begin{figure}[h!]
\rotatebox{-90}{\epsfig{file=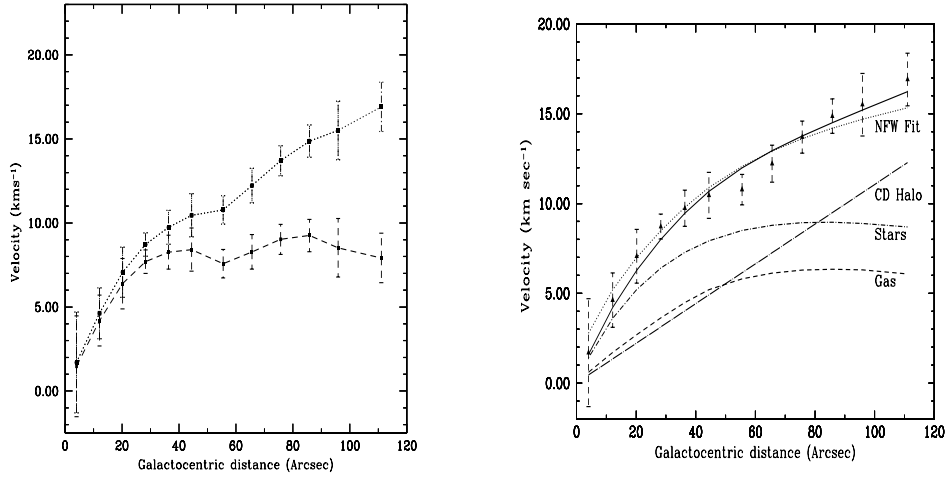,width=3.0in,height=5.5in}}
\caption{{\bf{[A]}}The hybrid rotation curve (dashes)  and the rotation curve
        after applying the asymmetric drift correction (dots).
{\bf{[B]}} Mass models for DDO210 using the corrected rotation curve.
         The points are the observed data. The total mass of gaseous
            disk (dashed line)  is $3.6\times10^6 M_\odot$ (including the contibution
	of primodial He).The stellar
          disk (short dash dot line) has
          $\Upsilon_B=3.4$, giving a stellar mass of $9.2 \times10^6 M_\odot$.
        The  best fit total rotation curve for the constant density
          halo model is shown as
          a solid line, while the contribution of the halo itself
           is shown as a  long dash dot line (the halo density is
           $\rho_0=29~\times10^{-3} M_\odot$ pc$^{-3}$).
         The best fit total rotation curve for an NFW type halo,
          using $\Upsilon_B=0.5$, c=5.0 and $v_{200}$=38.0~km s$^{-1}$
    is shown as a dotted line. See text for more details.
  }
\label{fig:ddo210_vrot}
\end{figure}

\subsection{DDO210}
\label{ssec:camb}

  DDO210 is the faintest known gas rich dwarf galaxy in our local group.
Karachentsev et al.(2002), based on the HST observations of the I magnitude of 
the tip of the red giant branch, estimated the distance to this galaxy to 
be 950$\pm$50 kpc.
Fig.~\ref{fig:ddo210_mom}[A] shows the integrated  HI column density image of 
DDO210  at 44$''\times 37''$
resolution overlayed on the optical DSS image. The HI isodensity contours are 
elongated in eastern and southern half of the galaxy, indicating a density enhancement
in these directions.The HI mass obtained from the integrated profile
(taking the distance to the galaxy to be 1.0 Mpc) is 2.8$\pm$0.3 $\times10^6$M$_\odot$ 
and the M$_{HI}$/L$_B$ ratio is found to be 1.0 in solar units.

 The velocity field of DDO210 derived from the moment analysis of 29$''\times 23''$
resolution data cube is shown in Fig.~\ref{fig:ddo210_mom}[B]. The velocity field is 
regular and a systematic velocity gradient is seen across the galaxy. 

 Fig.~\ref{fig:ddo210_vrot}[A] shows the derived rotation curve for DDO210 and the
``asymmetric drift" corrected rotation curve.  We find that the ``asymmetric
drift" corrected rotation curve of DDO210  can be well fit with  a
modified isothermal halo (with a central density $\rho_0 \sim 29\times10^
{-3}$ $M_\odot$ pc$^{-3}$). In the case of the NFW halo, we find that 
there are a range of parameters which provide acceptable fits e.g. (v$_{200}\sim$ 20 kms$^{-1}$, 
c $\sim$ 10) to (v$_{200}\sim$ 500 kms$^{-1}$, c$\sim$ 0.001); the halo parameters
could not be uniquely determined from the fit. However, the best fit value of the 
concentration parameter c, at any given v$_{200}$ was found to be consistently smaller 
than  the value predicted by numerical simulations for the $\Lambda$CDM universe (Bullock et al. 2001). 
Fig.~\ref{fig:ddo210_vrot}[B] shows the best fit mass models for DDO210 (see Begum \& Chengalur 2004 for details).

\begin{figure}[h!]
\rotatebox{0}{\epsfig{file=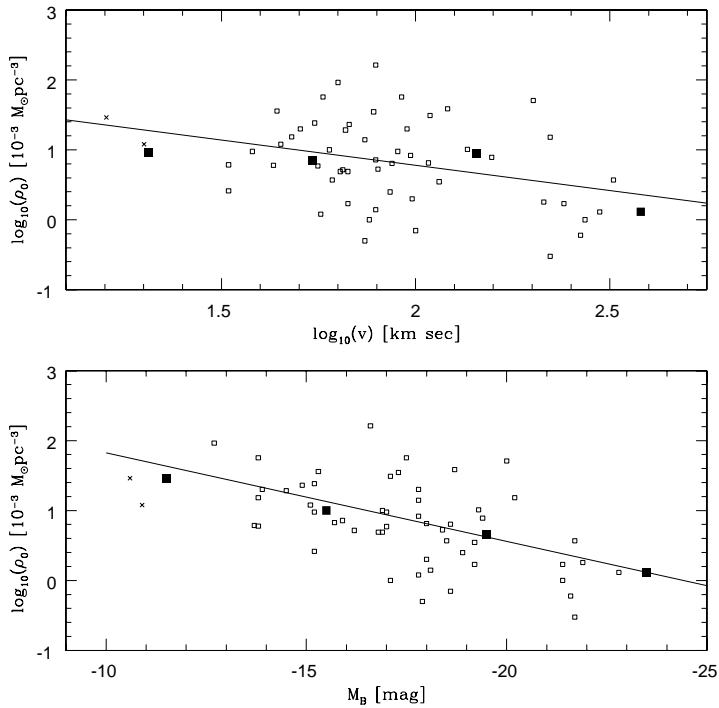,width=4.0in}}
\caption{Scatter plots of the central halo density against the
          circular velocity and the absolute blue magnitude . The data (empty squares)
          are from Verheijen~(1997), Broeils~(1992),
          C\^{o}t\'{e} et al.~(2000) and Swaters~(1999). The filled
          squares are the medians of the binned data, and the straight
          lines are the best fits to the data. Cam~B and DDO210 are
            shown  as crosses.
         }
\label{fig:dens}
\end{figure}

\section{Discussion}
\label{sec:result}

Fig.~\ref{fig:dens} shows  the core density of isothermal halo against circular
velocity and absolute blue magnitude for a sample of galaxies, spanning a range of
magnitudes from M$_B\sim-10.0$ mag to M$_B\sim-23.0$ mag. Cam B and DDO210 are also
shown in the figure, lying at the low luminosity end of the sample. 
 As can be seen in the figure,
there is a trend of increasing halo density with a decrease in circular velocity
and absolute magnitude, shown by a best fit line to the data (solid line),
although the correlation is very weak and noisy. Further, as a guide to an
eye, we have also binned the data and plotted the median value (solid points).
Binned data also shows a similar trend. Such a tread is expected in
hierarchical structure formations scenario (e.g. Navarro,
Frenk \& White 1997).

\end{document}